# A Schrödinger Equation for Quantum Universes

*Marco Cavaglià*[1,4], *Vittorio de Alfaro*[2,4] *and Alexandre T. Filippov*[3]

[1] Int. School for Advanced Studies, Via Beirut 2-4, I-34013, Trieste, Italy
[2] Dip. di Fisica Teorica dell'Università di Torino, Via Giuria 1, I-10125 Torino, Italy
[3] Joint Institute for Nuclear Research, R-141980 Dubna, Moscow Region, Russia
[4] Istituto Nazionale di Fisica Nucleare, Sezione di Torino, Italy

During the last years, many attemps have been done to formulate a quantum theory of gravity [1]. A well defined quantum theory of the gravitational field is very important when one discusses the birth of the Universe because at the Planck temperatures the classical theory is no longer valid. The present workshop is dedicated to the physics of the early Universe and so it is appropriate to discuss an important problem that arises when trying to give a quantum description of our Universe: the definition of time (see also in this volume the contributions by Mensky and Albrecht).

Naturally, up to now a well established quantum gravity theory does not exist, so general considerations about the nature of time are out of reach and only simple minisuperspace models can be investigated. Minisuperspace models are finite-dimensional systems and their quantization can be completed - at least in principle. For instance, in the case of a Friedmann-Robertson-Walker Universe filled with matter and radiation something can be said and a quantum Schrödinger equation can be written quantizing the system in according to the usual canonical formalism [2]. This model can have some relevance to the quantum cosmology program - the observed Universe is homogeneus and isotropic on the average - and its features will allow us to explore how one can recover the dynamical evolution in a quantum model of the Universe.

Classical General Relativity is a theory invariant with respect to time reparametrization [3]. So in minisuperspace models the extended Hamiltonian is of the form $H = l(t)\mathcal{H}(q,p)$ where $l(t)$ is a Lagrange multiplier that imposes the constraint $\mathcal{H} = 0$. When quantizing the model, in order to select the physical states of the system one has to impose this constraint on the wave function: $\mathcal{H}\Psi = 0$. In the latter equation there is no time parameter and thus no dynamical evolution. This is not surprising. The absence of dynamics arises because there are no external observers measuring the evolution of the Universe: time is a parameter internal to the system. Thus, how can we recover the dynamics or, in other words, how can we derive the Schrödinger equation which describes the evolution of the Universe? A possible way out to solve this puzzle is quantizing the classical model by imposing a gauge fixing condition. In doing that, one reduces the space phase to the physical degrees of freedom and so a dynamical Schrödinger equation in the remaining variables can be written [2]. Let us see this approach in detail.

Whenever a classical system is described by a theory with $N$ gauge constraints, $2N$ canonical degrees of freedom of the system are redundant [3]. Typical examples are the relativistic particle or the relativistic quantum oscillator. In our minisuperspace models, we have only one constraint - $\mathcal{H} = 0$ - due to time reparametrization. This redundancy can be eliminated by a gauge fixing condition, namely an (arbitrary) relation between the canonical variables and time: $\mathcal{F}(q,p;t) = 0$. $\mathcal{F}$ is chosen so that the Poisson bracket $\{\mathcal{F}, \mathcal{H}\}$ does not vanish (even weakly) and then the gauge fixing and the constraint determine the Lagrange multiplier as a function of time and canonical variables and allow to reduce the space phase to the physical degrees of freedom. In this way, a reduced effective Hamiltonian can be recovered and (in principle) its quantization can be performed: the remaining canonical variables corresponding to the physical degrees of freedom can be treated as operators on a Hilbert space of wave functions and a Schrödinger equation can be written. Let us emphasize that in the reduction quantization method the choice of time is performed classically. Different gauge fixings select different time choices, and so generate different effective Hamiltonians, i.e. different dynamical evolutions of the system.

Of course, practical and conceptual problems arise when one applies the reduction quantization method described above: gauge fixings often lead to a non unitary evolution and it is not clear whether different time choices produce identical effective theories. A closed Friedmann-Robertson-Walker Universe filled with radiation and matter fields is an appropriate laboratory to discuss the gauge fixing and explore the difficulties mentioned above. For this model the gravitational part of the Hamiltonian coincides essentially with the Hamiltonian of a harmonic oscillator. Moreover, when only radiation fields are present (for instance a Yang-Mills field or a conformal scalar field), the gravitational and matter Hamiltonians are separable. These features make possible to illustrate a different number of gauge fixings and time choices, with their corresponding characteristics. Let us see in detail some cases.

**Friedmann-Robertson-Walker Universe filled with radiation fields.** In this case the constraint is

$$\mathcal{H} = \frac{1}{2}p_a^2 + \frac{1}{2}a^2 - H_{rad}(q_i, p_i) = 0, \qquad (1)$$

where $a$ is the scale factor of the Universe and $p_a$ its conjugate momentum. A viable reduction can be obtained through the canonical transformation $P_a = p_a - a \operatorname{ctg} t$ and the gauge fixing condition $P_a = 0$. The effective Hamiltonian on the gauge shell is

$$H_{eff} = H_{rad} \qquad (2)$$

while the lapse function is fixed to be $N = a$. Thus, in the conformal gauge the quantum dynamical equation on the gauge shell takes the form

$$i\frac{\partial}{\partial t}\Psi(q_i;t) = H_{rad}\Psi(q_i;t). \qquad (3)$$

Eq. (3) is a dynamical equation and $\Psi$ represents, as usual, the correlation amplitude for the different components of matter at given $t$. Another interesting gauge is given by $p_a = t$. In this case the lapse function is fixed as $N = 1$ and so corresponds to the gauge of a comoving free falling observer. The effective Hamiltonian is

$$H_{eff} = \sqrt{2H_{rad} - t^2}. \tag{4}$$

The Schrödinger equation is

$$i\frac{\partial}{\partial t}\Psi(q_i;t) = \sqrt{2H_{rad} - t^2}\,\Psi(q_i;t). \tag{5}$$

So the quantum theory is non unitary because the classical Hamiltonian is real only in a finite range of $t$. At the moment is it not clear how to treat Hamiltonians like (5).

**Friedmann-Robertson-Walker Universe filled with a minimally coupled scalar field.** The presence of a minimally coupled scalar field is of great interest as it introduces a coupling to gravitation that induces inflation. The constraint is

$$\mathcal{H} = \frac{1}{2}p_a^2 + \frac{1}{2}a^2 - \frac{1}{2}\frac{p_\Phi^2}{a^2} - a^4 V(\Phi) - H_{rad} = 0. \tag{6}$$

Due to the structure of (6), only particular cases can be discussed. In the slow-rolling phase, we can disregard the kinetic term. In this case, choosing for instance the gauge fixing $a = t/\sqrt{2}$, the effective Hamiltonian becomes

$$H_{eff} = \sqrt{H_{rad} + t^2(V(\phi_0)t^2 - 1)/4} \tag{7}$$

As for (5), when $H_{rad} = 0$ the Hamiltonian is not real for $t < V(\Phi)^{-1/2}$, i.e. for the classical forbidden region. In the opposite case, when the potential term and $H_{rad}$ are negligible, an interesting gauge is $p_a = a \sinh(2t)$. With this choice, the effective Hamiltonian is

$$H_{eff} = \pm p_\phi \tag{8}$$

and the Schrödinger equation is

$$\left(\frac{\partial}{\partial t} - (\pm)\frac{\partial}{\partial \phi}\right)\Psi = 0 \tag{9}$$

which has the general solution

$$\Psi = f(\phi \pm t). \tag{10}$$

In conclusion, the reduction quantization method can be successfully implemented for some minisuperspace models to obtain a quantum dynamical Schrödinger equation. Of course, a general method that can work for any model does not exist and much must be learned about the nature of time in quantum gravity. However, minisuperspace models can help us to shed light on the charming issue of the birth of the Universe.

**Acknowledgements.** It is a pleasure to thank the organizers of the workshop and in particular Franco Occhionero and Anna Restante for their warm hospitality during the workshop.